\documentclass{PoS}
\usepackage{bbm,slashed,amsmath,amssymb}
\newcommand{\braket}[1]{\left\langle #1 \right\rangle}
\DeclareMathOperator{\re}{Re}
\DeclareMathOperator{\im}{Im}
\newcommand{\fieldInf}[0]{\Psi}
\newcommand{\fieldFinite}[0]{\phi}

\title{Towards the large volume limit -- A method for lattice QCD + QED simulations}
\ShortTitle{Towards the large volume limit}

\author{\speaker{Christoph Lehner}\\
        Physics Department, Brookhaven National Laboratory, Upton, NY 11973, USA\\
        E-mail: \email{clehner@quark.phy.bnl.gov}}

\author{Taku Izubuchi\\
        RIKEN-BNL Research Center, Brookhaven National Laboratory, Upton, NY 11973, USA\\
        Physics Department, Brookhaven National Laboratory, Upton, NY 11973, USA\\
        E-mail: \email{izubuchi@quark.phy.bnl.gov}}

      \abstract{We present a method to couple finite-volume QCD to
        infinite-volume QED by an appropriate twist-averaging
        procedure.  We demonstrate the prescription numerically for
        the leading-order hadronic contribution to the anomalous
        magnetic moment of the muon and the electro-magnetic pion mass
        splitting.}

\FullConference{The 32nd International Symposium on Lattice Field Theory,\\
		23-28 June, 2014\\
		Columbia University New York, NY}

\begin{document}

\section{Motivation}
The long-distance nature of QED poses a substantial challenge when
including QED interactions in lattice QCD simulations.  While it is
possible to account analytically for the large power-law corrections
that are introduced by the QED interaction in finite-volume QCD+QED
simulations, see, e.g., \cite{BMW, DAVOUDI}, such a setup requires the
use of lattice ensembles that are much larger than necessary for a
pure QCD simulation.  An alternative procedure that explicitly couples
regular QCD simulations in a finite volume (QCD$_V$) to QED in
infinite volume (QED$_\infty$) is presented here.

In the following we first introduce a specific
prescription to put valence fermions and photons in infinite volume
and then demonstrate two versions of the method in a numerical
application to the leading-order hadronic vacuum polarization (HVP).
We present results for electro-magnetic mass splittings and give
an outlook to general QCD$_V$ + QED$_\infty$ simulations.

\section{The setup}
\begin{figure}[th]
  \centering
  \includegraphics[width=9.5cm]{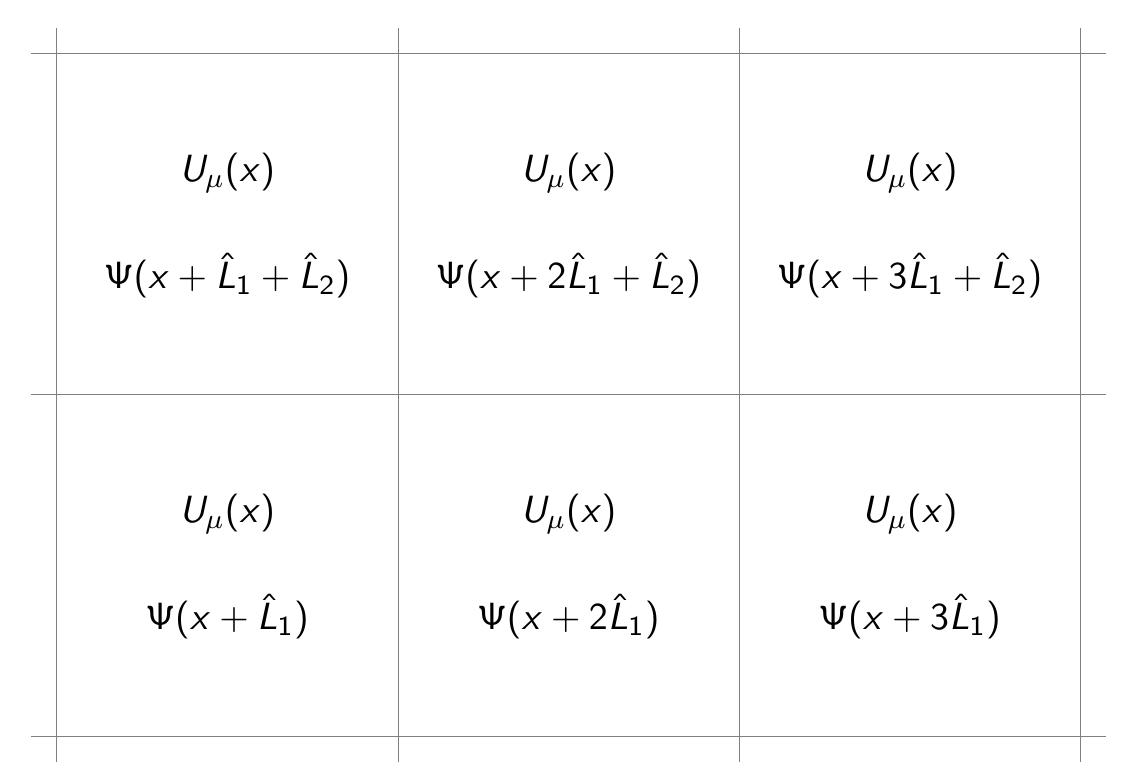}  
  \caption{Valence fermions $\fieldInf$ living on
    a repeated gluon background $U_\mu$ with periodicity $L_1$, $L_2$ and
    vectors $\hat{L}_1 = ( L_1, 0)$, $\hat{L}_2 = (0,L_2)$.}
  \label{fig:setup}
\end{figure}

Let us consider a box of QCD gauge fields $U_\mu$ periodic in a volume
$V$.  We furthermore imagine that we repeat the QCD gauge field in
each direction an infinite number of times and couple an
infinite-volume valence fermion $\fieldInf$ to the repeated gauge
background as illustrated in Fig.~\ref{fig:setup}.  The
infinite-volume fermion is in turn coupled to infinite-volume photons
$A_\mu$ via $A_\mu(x) V_\mu(x)$, where $V_\mu$ is a vector current of
fields $\fieldInf$.  One may imagine that such a setup with photons
living in infinite volume has largely suppressed QED
finite-volume errors since there are no mirror charges and the
finite-volume sum over photon momenta is explicitly replaced by the
infinite-volume integral.

If restricted to a finite number of QCD gauge field copies, this setup
is identical to solving the Dirac operator on the repeated QCD gauge
background but with QED gauge fields living in the larger volume.  In
the following we present a prescription to generate the
infinite-volume (or large-volume) setup stochastically.

Let $\fieldFinite^\theta$ be a fermion field defined in finite volume with
twisted boundary conditions
\begin{align*}
  \fieldFinite^\theta_{x+L_\mu \hat{\mu}} = e^{i\theta_\mu}\fieldFinite^\theta_x \,,
\end{align*}
where $\mu=\{x,y,z,t\}$, $L_\mu$ is the size of the QCD box in
direction $\mu$, $\hat{\mu}$ is the unit vector in direction $\mu$,
and $\theta_\mu$ is the twist angle in direction $\mu$.  In analogy to
Bloch's theorem, the symmetry under translations by $L_\mu \hat{\mu}$
then yields
\begin{align}\label{eqn:master}
  \braket{\fieldInf_{x + n L} \bar\fieldInf_{y + m L}} = \int_0^{2\pi} \frac{d^4\theta}{(2\pi)^4} e^{i\theta (n - m)}
  \braket{\fieldFinite^\theta_x \bar\fieldFinite^\theta_y } \,,
\end{align}
where the $\langle\cdot\rangle$ denotes the fermionic contraction in a
fixed background gauge field $U_\mu(x)$, $L=\sum_\mu \hat{\mu} L_\mu$,
and $x,y,n,m,\theta$ are four-vectors.  For a finite number $N_\mu$ of
QCD gauge field copies in direction $\mu$, the integral $\int_0^{2\pi}
\frac{d\theta_\mu}{2\pi}$ is replaced by the sum $(1 /
N_\mu)\sum_{\theta_\mu=0,2\pi/N_\mu,\ldots,2\pi(N_\mu-1)/N_\mu}$.  For
$N_\mu = 2$ this reduces to the well-known periodic-plus-antiperiodic
trick.  The twist-average of Eq.~\eqref{eqn:master} may be performed
stochastically and can be used to generate the infinite-volume photon
momentum integrals, see Sec.~\ref{sec:stochintphoton}.

The reduction of finite-volume errors through averaging of solutions
with different twists is well-known in the study of metallic systems
\cite{metallic} but was also recently investigated in the study of
two-baryon systems \cite{briceno}.

For a general observable, we propose the following prescription:
\begin{enumerate}
\item Use the infinite-volume fields $\fieldInf$ for all operators and sources,
\item perform the Wick contractions, and finally
\item use Eq.~\eqref{eqn:master} to write the result in terms of
  integrals over twists involving only Dirac inversions of the
  finite-volume theory.
\end{enumerate}

The positions of the vertices in the resulting expression can then be
summed over the infinite volume.  For quark-connected diagrams such a
sum can be performed na\"{i}vely as described below.  For
quark-disconnected contributions one needs to restrict the position of
separate quark loops to the same QCD box.\footnote{We would like to
  thank Luchang Jin for bringing this point to our attention.}  Such a
restriction can be readily implemented in the prescription discussed below.


The setup described in this section works well with
multi-source methods such as AMA \cite{AMA}.  In the remainder of this
manuscript the numerical data uses an AMA setup with only one twist
vector per source (such that the number of twist vectors that are
averaged coincides with the number of sources).  Note that 
one may be able to re-use zero-twist eigenvectors by employing solvers
such as \cite{HDCG} that use blocked eigenvectors.

\section{The muon hadronic vacuum polarization}
\begin{figure}[th]
  \centering
  \includegraphics[width=5cm]{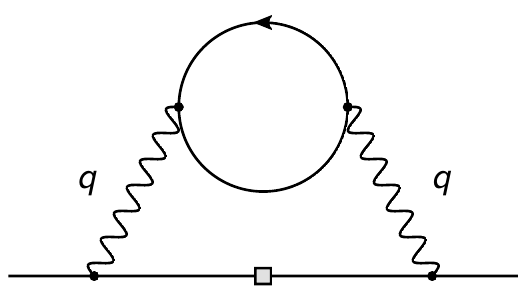}
  \caption{The leading-order hadronic vacuum polarization diagram.  The lower fermion line is the muon.}
  \label{fig:hvp}
\end{figure}

The leading-order hadronic vacuum polarization diagram, shown in Fig.~\ref{fig:hvp},
is conveniently evaluated following \cite{Blum:2002} which expresses the full diagram
as
\begin{align}\label{eqn:amu}
  a_\mu = \int_0^\infty d(q^2) F(q^2) \hat{\Pi}(q^2)
\end{align}
with function $F(q^2)$ defined in \cite{Blum:2002} capturing the photon
and muon components of the diagram, $\hat{\Pi}(q^2) = \Pi(q^2) -
\Pi(0)$, $\Pi_{\mu\nu}(q) = (q_\mu q_\nu - q^2 g_{\mu\nu}) \Pi(q^2)$.
Furthermore
$\Pi_{\mu\nu}(q) = \sum_x e^{iqx} \Pi_{\mu\nu}(x)$ and
$\Pi_{\mu\nu}(x)=\braket{ V_\mu(x) V_\nu(0) }$ with vector current
$V_\mu$.  In the following numerical examples, we use a conserved
vector current at position $x$ and a local vector current at the origin.

Equation \eqref{eqn:amu} follows the prescription that we
outlined above.  We couple the QCD fermion loop to an infinite-volume
photon whose propagators are included in the function $F(q^2)$.  By
computing $\Pi_{\mu\nu}$ defined in terms of the infinite-volume
fields $\fieldInf$, we have access to a continuum of photon momenta
$q$.  The most significant simplification of this specific example is
that we have reduced the four-dimensional photon momentum integral
analytically to a one-dimensional integral over $q^2$.  In this way, we
can test our prescription in the case of twisting in just a single
direction.  In a general QCD+QED problem, we will evaluate
four-dimensional photon-momentum integrals and appropriately included
non-zero twist angles in all four directions.

In the remainder of this section we first explore a na\"{i}ve
implementation of the twist-averaging idea to evaluate $a_\mu$.  Let
$C_{\mu\nu}(t) = \sum_{\vec{x}} \Pi_{\mu\nu}(x_0=t,\vec{x})$, $C(t) =
C_{\mu\mu}(t)$ for $\mu = x,y,z$.  Then we can write down an estimator
for $\hat{\Pi}(q^2)$ that configuration-by-configuration satisfies
$\hat{\Pi}(q^2 = 0) = 0$ and thereby avoids a statistical noise
problem for small momenta.  Starting from $\braket{
  \Pi_{\mu\nu}(q^2=0) } =0$ and $ \braket{ \im \Pi_{\mu\nu}(q^2) } =
0$ we arrive at
\begin{align}
  \braket{ \hat{\Pi}(q^2) } = \braket{ \sum_t \re \left(\frac{\exp( i q t ) - 1}{q^2} + \frac12 t^2 \right) \re C_{\mu\mu}(t) }
\end{align}
which is a minimal modification of Eq.~(81) of \cite{BerneckerMeyer2012}.  This expression includes an explicit double-subtraction of
$\Pi_{\mu\nu}(q^2=0)$ and $\Pi(q^2=0)$.

More specifically, we compute
\begin{center}
  \includegraphics[scale=0.8]{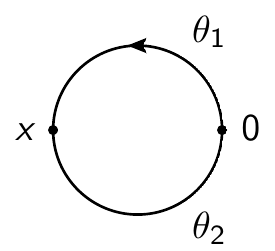}
\end{center}
averaging over independent temporal twist angles $\theta_1$ and
$\theta_2$ for the two propagator lines.  All spatial twists are zero
such that our setup corresponds to an infinite time direction.  We
inject momentum only in time direction and cut the sum over time-slices
at sufficiently large distance such that the contribution of the
remainder can be ignored within the statistical precision.  In
Sec.~\ref{sec:stochintphoton}, we present a method that does not
require such a cut.

\begin{figure}[th]
    \centering
    \includegraphics[height=5cm,page=3]{figs/009-corr}~~~~
    \includegraphics[height=5cm,page=1]{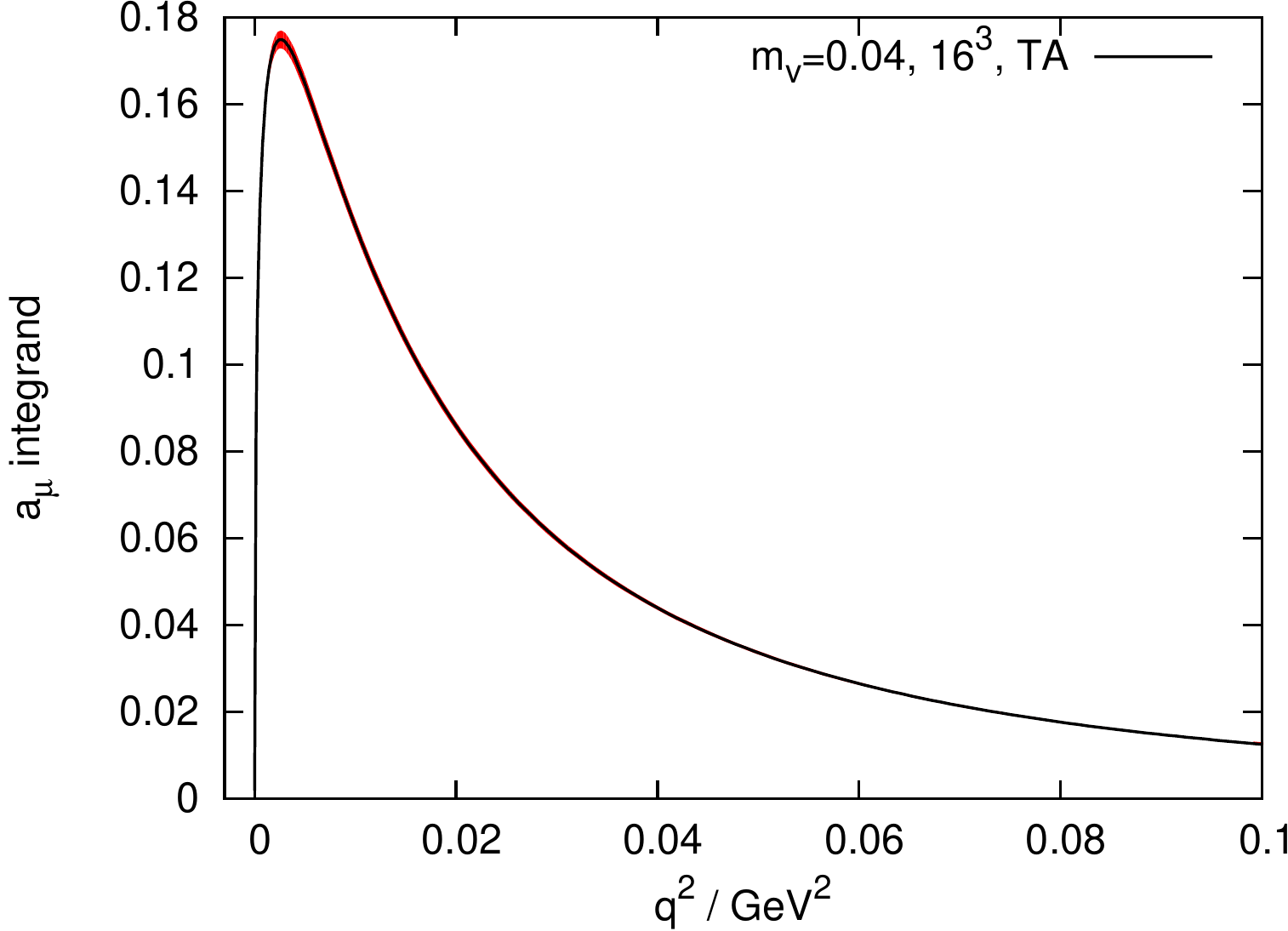}
    \caption{Strange quark contribution on RBC/UKQCD's $16^3$ and
      $24^3$ ensembles with $a^{-1}\approx 1.73$ GeV, $m_l=0.01$
      ($m_\pi \approx 422 $ MeV), and $m_s=0.04$.  The $16^3$ ($24^3$)
      measurement was performed using 2 (1) exact and 32 (16) inexact
      sources on 60 (78) configurations.  The left figure shows the
      times-lice $t$ contribution to $\hat{\Pi}(k=m_\mu)$, where
      $m_\mu$ is the physical muon mass.  The right figure shows the
      integrand of $a_\mu$ as a function of $q^2$.  Note
      that the stochastic noise is well-behaved for small
      momenta.} 
    \label{fig:fvtest}
  \end{figure}

Figure \ref{fig:fvtest} shows a numerical comparison of results using
RBC/UKQCD's $16^3$ and $24^3$ ensembles that only differ in their
respective physical volume.  This allows for a test of remnant
finite-volume errors introduced by the sea sector.  The integral over
$q^2$ is performed using both the Trapezoidal and Simpson's rule,
choosing a step-size such that their difference is below $1/100$ of the
statistical error.  We find that the relative error of $a_\mu$ is
consistent with the almost $q^2$-independent integrand uncertainty
shown in Fig.~\ref{fig:resultamu}.


\newcommand{\N}{\mathbbm{N}}

\section{Stochastic integration of photon momenta}\label{sec:stochintphoton}
In this section we demonstrate an efficient method to use the twist-averaging procedure
to sample over the photon momenta.  A wise choice of sampling weight, i.e., the use
of importance-sampling techniques can yield a substantial benefit.  The following
discussion is explicitly given in one dimension but all expressions and methods
translate in a straightforward way to the more general four-dimensional case.

We continue the discussion of the HVP diagram to illustrate the method.  The full diagram with lattice regulator can be written as
\begin{align}\label{eqn:stochsamp}
 \int_{-\pi}^\pi dk \: \Gamma^{\mu\nu}(k) \Pi_{\mu\nu}(k) = \int_{-\pi}^\pi dk \: \Gamma^{\mu\nu}(k) \int_0^{2\pi} \frac{d\theta_1}{2\pi}\frac{d\theta_2}{2\pi}\sum_{x \in \{0,\ldots,L-1\}, n \in \N} e^{i k (n L + x) + i n (\theta_1 - \theta_2)} C_{\mu\nu}(x,\theta_1,\theta_2) 
\end{align}
for an appropriately defined $\Gamma^{\mu\nu}$.  Poisson's summation formula yields
\begin{align}
 \int_{-\pi}^\pi dk \: \Gamma^{\mu\nu}(k) \int_0^{2\pi} \frac{d\theta_1}{2\pi}\frac{d\theta_2}{2\pi}\sum_{x \in \{0,\ldots,L-1\}} e^{i k x} \hat\delta(k - (\theta_2-\theta_1)/L) C_{\mu\nu}(x,\theta_1,\theta_2)
\end{align}
with $ \hat\delta(k) = \frac{2\pi}{L} \sum_{n \in \N} \delta(k + 2\pi n / L) $.  By writing
$\int_{-\pi}^\pi dk\: g(k) = \frac1L \sum_{n=0,\ldots,L-1} \int_0^{2\pi} d\theta_k g(2\pi n/L + \theta_k / L)$
and performing the integral over $\theta_k$, we obtain a method to stochastically sample the photon momenta
through the random choice of twist angles.

\begin{figure}[th]
  \centering    
  \includegraphics[width=6cm,page=6]{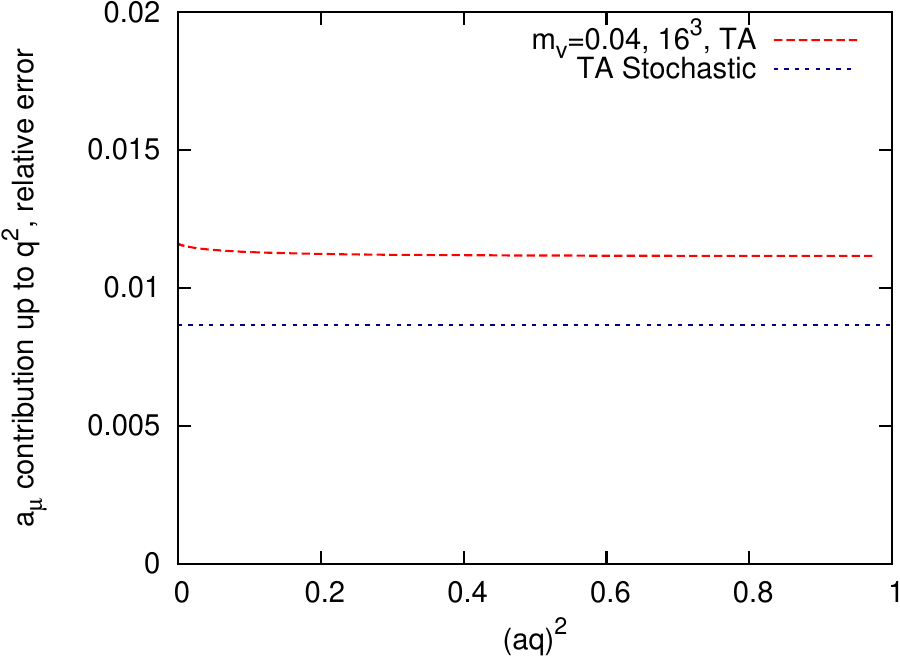}~~~~~~
  \includegraphics[width=6cm,page=4]{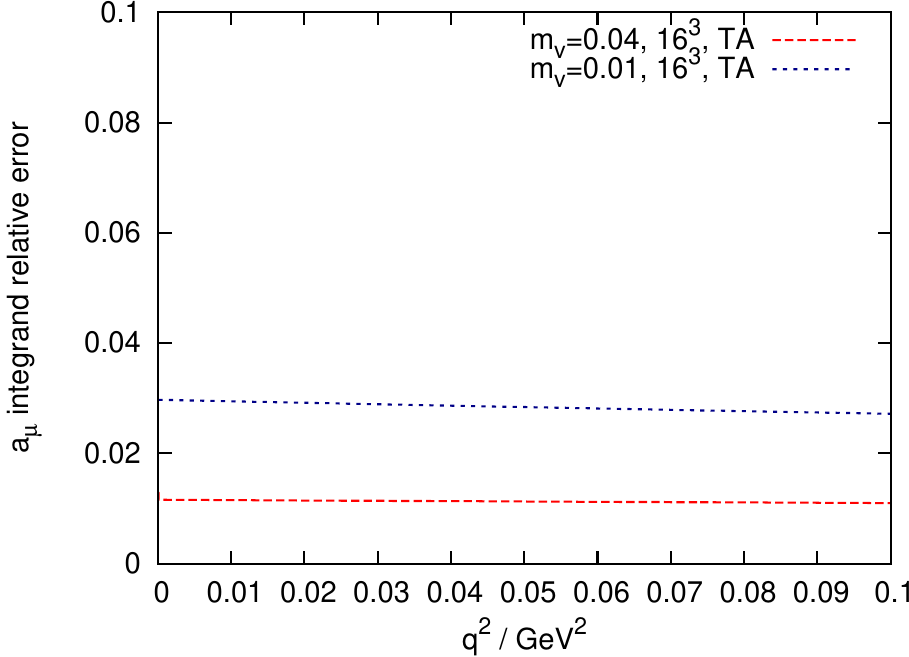}
  \caption{Relative errors of $a_\mu$ integrated up to maximal
    momentum $q^2$ (left) as well as the integrand of $a_\mu$ for
    $q^2$ (right) for the methods described in this manuscript.}
  \label{fig:resultamu}
\end{figure}

In Fig.~\ref{fig:resultamu} we show the resulting error for this method using the same configurations and sources as used in Fig.~\ref{fig:fvtest}.
We now, however, use importance sampling for the random twist angles such that they follow the probability distribution induced by
$F(q^2)$ of Eq.~\eqref{eqn:amu}.  We observe a slight reduction of statistical error, which may demonstrate the benefit of the importance
sampling strategy.

The sum over $L$-translations in Eq.~\eqref{eqn:stochsamp} generates conservation of the small-momentum component
$\theta_k = \theta_2 - \theta_1$.  This leads to an intuitive pictorial prescription in momentum space
\begin{center}
  \includegraphics[scale=0.7]{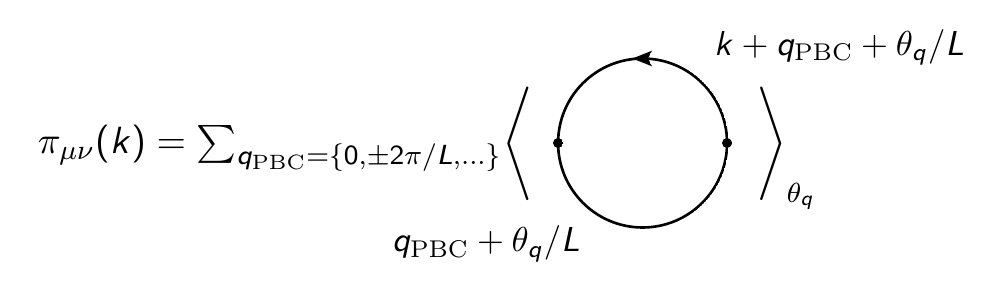}  
\end{center}
with average over twist-angle $\theta_q$ and sum over finite-volume periodic momenta $q_{\rm PBC}$.

\section{QED mass splittings}\label{sec:emmass}
A potentially interesting application of the ideas explained above is the computation of QED mass splittings.
For a general meson two point function, one needs to evaluate (ignoring quark-disconnected diagrams for now)
\vspace{-0.12cm}
\begin{center}
  \hskip-4cm\includegraphics[scale=0.6]{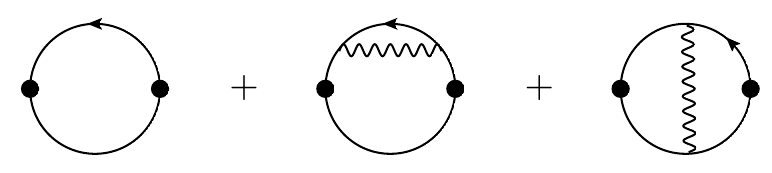} \vskip-0.9cm \hskip4cm $= C_0(t) + \alpha C_1(t)$     
\end{center}
\vspace{0.12cm}
for which the mass-shifts could be obtained from effective mass plots
following
\begin{align*}
  m_{\rm eff}(t) = m_{\rm eff,0}(t) + \alpha m_{\rm eff,1}(t) \,, \qquad
  m_{\rm eff,1}(t) = \frac{C_1(t)}{C_0(t)} - \frac{C_1(t+1)}{C_0(t+1)} \,.
\end{align*}

In this general case a four-dimensional photon momentum integral and hence
a four-dimensional twist-averaging procedure is necessary.  Using an appropriate
importance-sampling of twist angles, the methods outlined above should be applicable
in a straightforward manner.  The photon propagator could, e.g., be implemented stochastically
in a sequential source setup.

\begin{figure}[th]
    \centering
    \includegraphics[width=6cm,page=3]{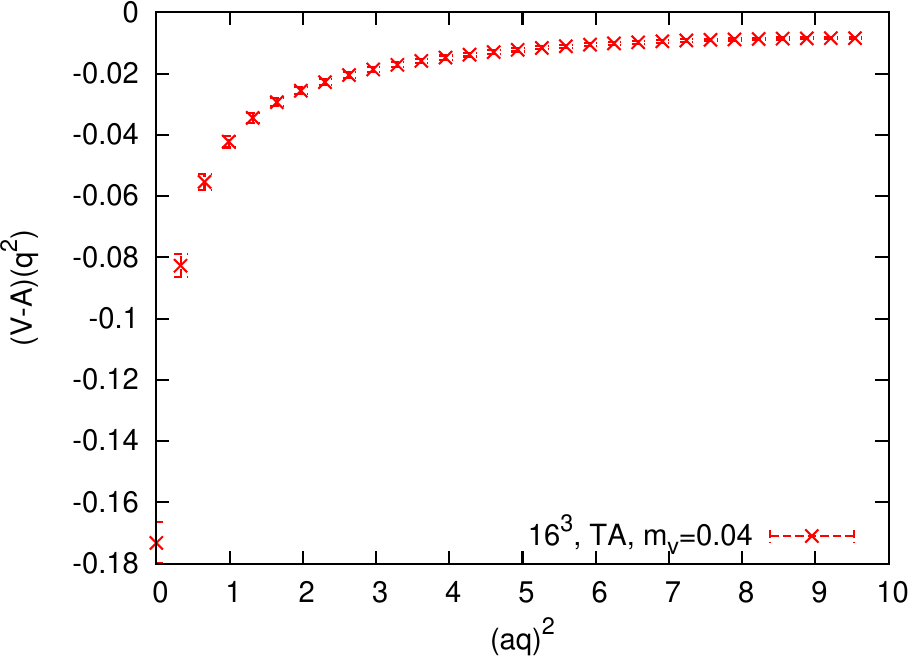}    
    \caption{V-A contribution for the strange quark loop.  At large $q^2$ there is a gap due to the heavy mass such that the integral does not converge.
    The stochastic integration method of the photon momentum with $(aq)^2<1$ cut yields an $\approx 4\%$ statistical error.}
    \label{fig:VmA}
  \end{figure}

For a first test that allows us to re-use the HVP measurements, we use current algebra and soft pion theorems \cite{softpion}
and compute
\begin{center}
  \includegraphics[scale=0.8]{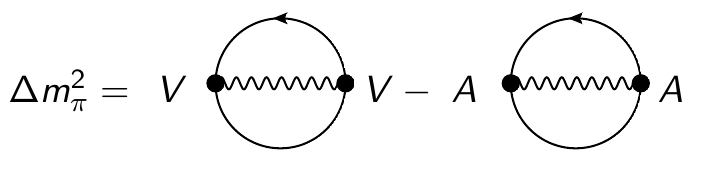}     \,.
\end{center}
Results of a quick numerical test are shown in Fig.~\ref{fig:VmA}.

\vspace{-0.5cm}
\section{Conclusion}
\vspace{-0.2cm}
We have proposed a prescription to put valence fermions and photons in infinite volume.  We
expect a substantial reduction of finite-volume errors in lattice QCD+QED simulations in this
setup such that regular lattice QCD boxes could be used for combined QCD+QED simulations.
We have discussed both a na\"{i}ve implementation of the idea and a refined strategy that samples
over photon momenta stochastically.  In the latter method an importance-sampling strategy seems
promising.

In the future, the methods presented here will be tested for realistic pion masses in the context
of QED mass splittings, QED corrections to decay constants, as well as finite-volume corrections to
the hadronic contributions to $(g-2)_\mu$, in particular the light-by-light contribution.

{\bf Acknowledgements} We thank our colleagues of the RBC and UKQCD
collaborations and in particular Tom Blum, Norman Christ, Luchang Jin,
and Chulwoo Jung for fruitful discussions.  T.I and C.L are supported
in part by US DOE Contract \#AC-02-98CH10886(BNL). T.I is also
supported by Grants-in-Aid for Scientific Research \#26400261.

\end{document}